\documentclass[sigconf]{acmart}

\usepackage{diagbox}
\usepackage{subfig}
\usepackage{multirow}
\usepackage{enumitem}
\usepackage{soul}
\usepackage[skip=3pt]{caption}

\acmJournal{TOG}
\copyrightyear{2022}
\acmYear{2022}
\setcopyright{acmcopyright}
\acmConference[SIGGRAPH '22 Conference Proceedings]{Special Interest Group on Computer Graphics and Interactive Techniques Conference Proceedings}{August 7--11, 2022}{Vancouver, BC, Canada}
\acmBooktitle{Special Interest Group on Computer Graphics and Interactive Techniques Conference Proceedings (SIGGRAPH '22 Conference Proceedings), August 7--11, 2022, Vancouver, BC, Canada}
\acmPrice{15.00}
\acmDOI{10.1145/3528233.3530733}
\acmISBN{978-1-4503-9337-9/22/08}

\begin{document}

\author{Yiwei Hu}
\affiliation{
	\institution{Yale University}
	\city{New Haven}
	\state{CT}
	\country{USA}
}
\affiliation{
	\institution{Adobe Research}
	\city{San Jose}
	\state{CA}
	\country{USA}
}
\email{yiwei.hu@yale.edu}

\author{Paul Guerrero}
\affiliation{
	\institution{Adobe Research}
	\city{London}
	\country{UK}
}
\email{guerrero@adobe.com}

\author{Miloš Hašan}
\affiliation{
	\institution{Adobe Research}
	\city{San Jose}
	\state{CA}
	\country{USA}
}
\email{mihasan@adobe.com}

\author{Holly Rushmeier}
\affiliation{
	\institution{Yale University}
	\city{New Haven}
	\state{CT}
	\country{USA}
}
\email{holly.rushmeier@yale.edu}

\author{Valentin Deschaintre}
\affiliation{
	\institution{Adobe Research}
	\city{London}
	\country{UK}
}
\email{deschain@adobe.com}

\renewcommand{\shortauthors}{Hu et al.}

\title{Node Graph Optimization Using Differentiable Proxies}

\begin{abstract}
Graph-based procedural materials are ubiquitous in 
content production industries. Procedural models  allow the creation of photo-realistic materials with parametric control for  flexible editing of appearance. However, designing a specific material is a time-consuming process in terms of building a model and fine-tuning parameters. Previous work \cite{Shi20, hu2022} introduced material graph optimization frameworks for matching target material samples. However, these previous methods were limited to optimizing differentiable functions in the graphs. 
In this paper, we propose a fully differentiable framework which enables end-to-end gradient-based optimization of material graphs, even if some functions of the graph are non-differentiable. We leverage the Differentiable Proxy, a differentiable approximator of a non-differentiable black-box function. We use our framework to match structure and appearance of an output material to a target material, through a multi-stage differentiable optimization. Differentiable Proxies offer a more general optimization solution to material appearance matching than previous work.
\end{abstract}



\begin{teaserfigure}
	\centering
	\addtolength{\tabcolsep}{-4pt}
	\begin{tabular}{ccccccccc}
	    Init (Proc.) & MATch (Proc.) & \textbf{Ours (Proc.)} & Target (Photo) &          & Init (Proc.) & MATch (Proc.) & \textbf{Ours (Proc.)} & Target (Photo)\\
		\includegraphics[width=0.12\textwidth]{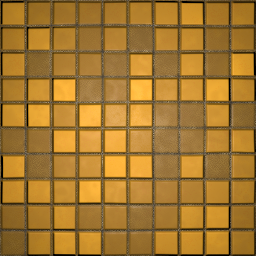} &
		\includegraphics[width=0.12\textwidth]{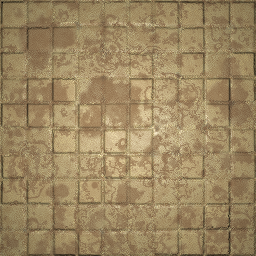} &
		\includegraphics[width=0.12\textwidth]{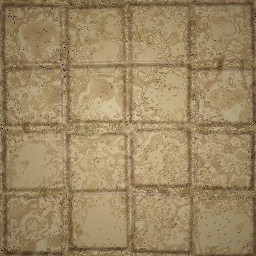} &
		\includegraphics[width=0.12\textwidth]{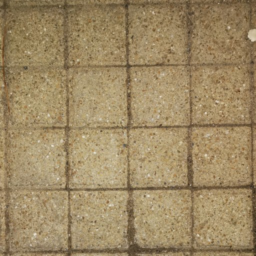} &          &
		\includegraphics[width=0.12\textwidth]{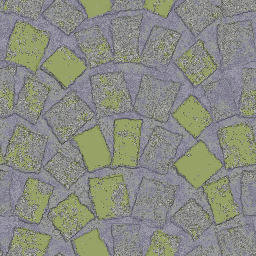} &
		\includegraphics[width=0.12\textwidth]{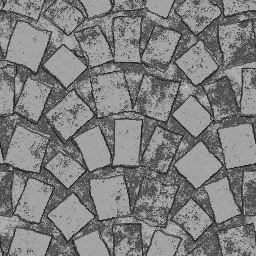} &
		\includegraphics[width=0.12\textwidth]{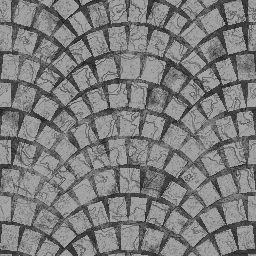} &
		\includegraphics[width=0.12\textwidth]{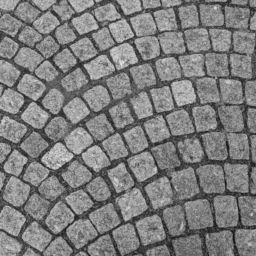} \\
		\includegraphics[width=0.12\textwidth]{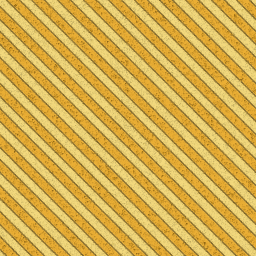} &
		\includegraphics[width=0.12\textwidth]{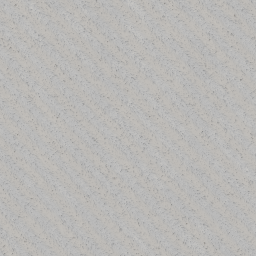} &
		\includegraphics[width=0.12\textwidth]{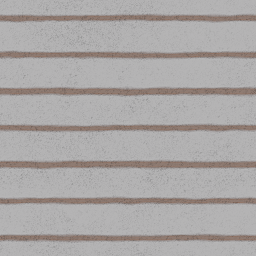} &
		\includegraphics[width=0.12\textwidth]{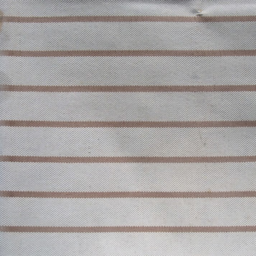} &          &
		\includegraphics[width=0.12\textwidth]{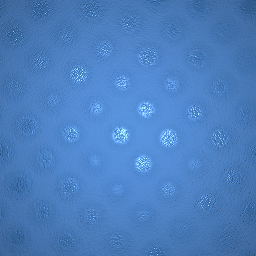} &
		\includegraphics[width=0.12\textwidth]{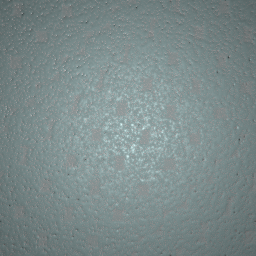} &
		\includegraphics[width=0.12\textwidth]{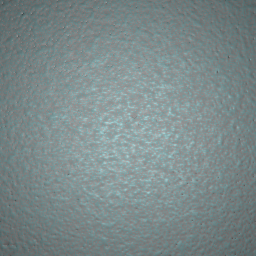} &
		\includegraphics[width=0.12\textwidth]{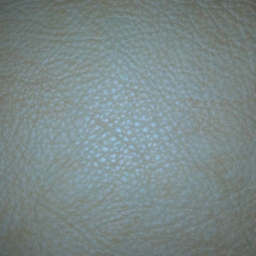} \\
	    \includegraphics[width=0.12\textwidth]{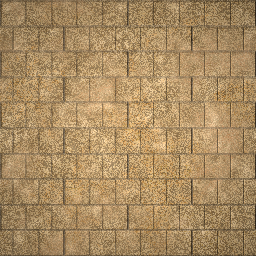} &
	    \includegraphics[width=0.12\textwidth]{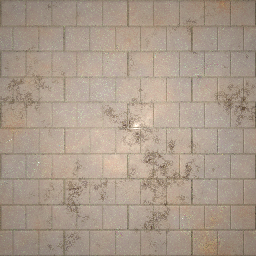} &
		\includegraphics[width=0.12\textwidth]{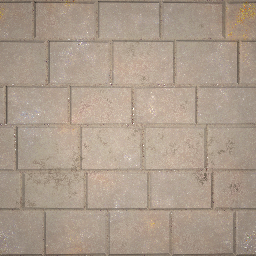} &
		\includegraphics[width=0.12\textwidth]{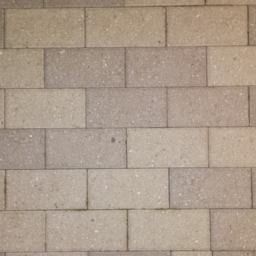} &          &
		\includegraphics[width=0.12\textwidth]{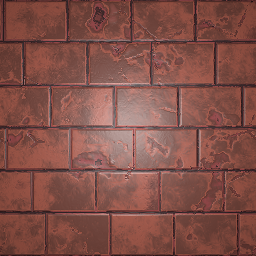} &
		\includegraphics[width=0.12\textwidth]{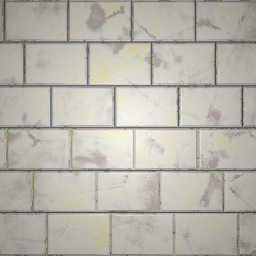} &
		\includegraphics[width=0.12\textwidth]{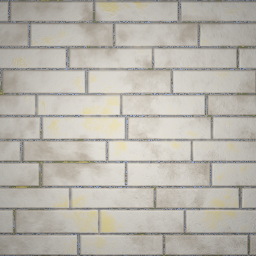} &
		\includegraphics[width=0.12\textwidth]{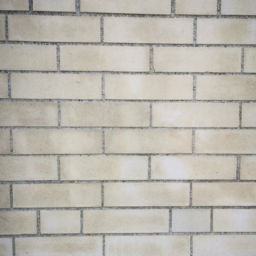} \\
	\end{tabular}
\caption{Given a user or classifier provided procedural graph, our method enables and performs end-to-end optimization of graph parameters towards the photograph of a surface. We show here results of our method, against previous work (MATch~\cite{Shi20}). In particular, compared to previous work, we enable gradient-based optimization of structure and scale of procedural materials. Images which are renderings of procedural models are marked with (Proc.)}
\label{fig:newTeaser}
\Description{Teaser and optimization results on real photos}
\end{teaserfigure}

\maketitle

\section{Introduction}
Virtual environment creation relies on a number of artist-created assets. Geometries, lighting, and materials constitute the core of complex environments in movies, video games, or architectural design. Graph-based modeling is commonly used in industry to design sophisticated effects. In this paper, we focus on graph-based procedural representations of materials, e.g., Spatially-varying Bidirectional Reflectance Distribution Functions (SVBRDFs), using parametric operators organized as a computational graph.

SVBRDFs can be represented by pixel maps encoding spatially varying material parameters of an analytical material model (e.g.,  Cook-Torrance~\shortcite{cook-torrance}). Pixel map representations are orders of magnitude more compact than tabulated BRDFs, but are limited to a fixed resolution and require inconvenient per-pixel operations to edit. Procedural graph models, on the other hand, provide parametric flexible control of materials and allow the generation of arbitrary-resolution material maps. The main challenge of procedural materials is the time-consuming modeling process~\cite{SubstanceDes}. To achieve a specific material appearance, trained artists need to carefully design the graph and hand-tune a number of parameters. Previous work ~\cite{hu2019, Shi20, hu2022} focused on reducing the design effort required by using various inverse modeling frameworks, but with limitations e.g., only considering parameters in differentiable nodes in a graph. In this paper, we propose using Differentiable Proxies for non-differentiable nodes to allow end-to-end optimization of entire material graphs.

\begin{figure}
    \centering
    \includegraphics[width=0.47\textwidth] {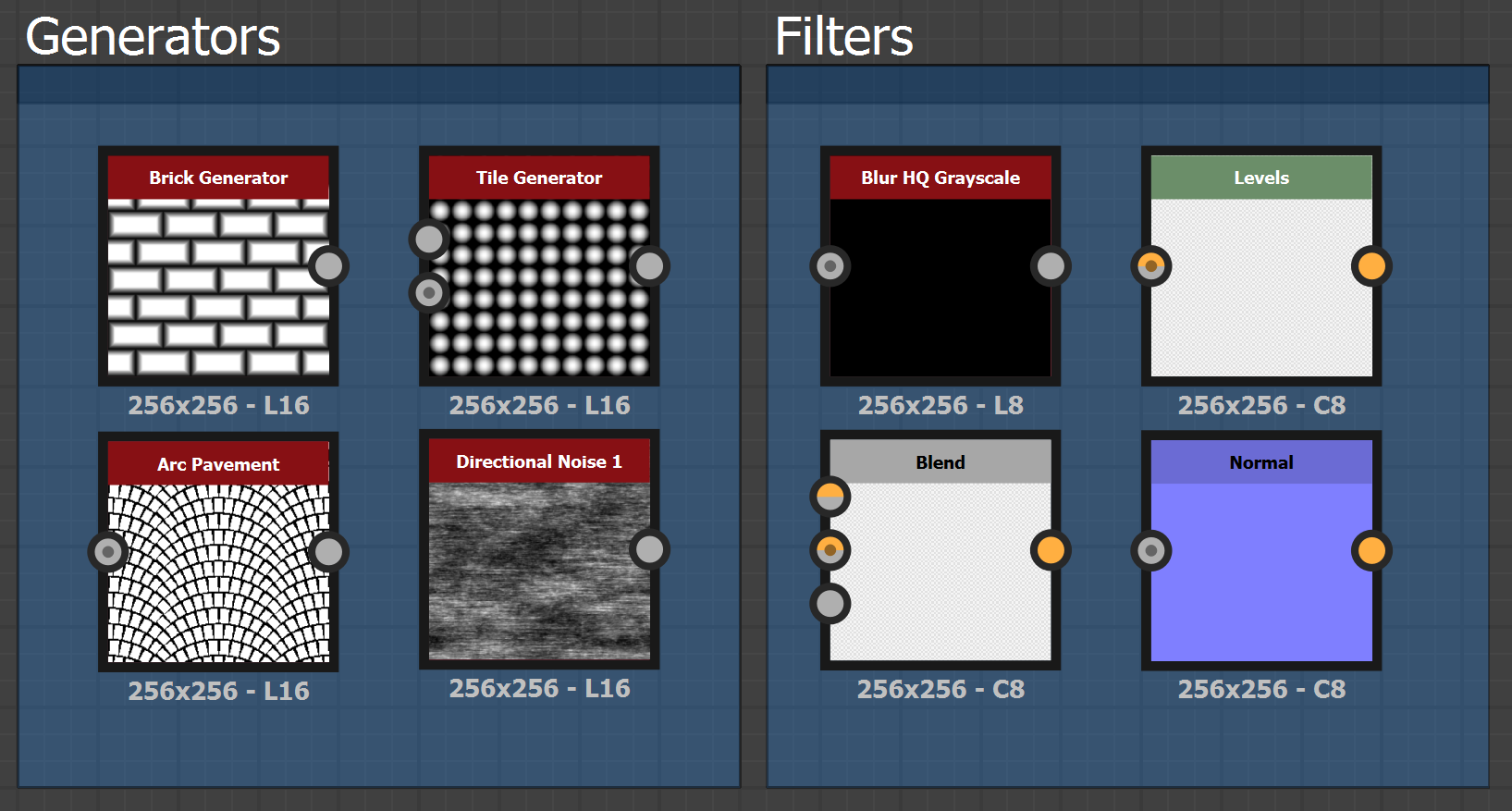}
    \caption{Typical generators and filters used in material graphs such as Substance Graphs where generators model elemental patterns, while filters are mostly smooth functions, modifying the input image appearance. }
    \label{fig:nodes}
    \Description{Introduction to node types}
\end{figure}

Nodes in a material graph can typically be classified into two types~\cite{SubstanceDes, Shi20, hu2022}: 1) \emph{Generators}, generating patterns and noises which usually require specifying discrete parameters; 2) \emph{Filters}, which are mostly smooth functions manipulating the generated patterns to reach the envisioned appearance. Fig. \ref{fig:nodes} shows a few commonly used generators and filters. 
In previous work, Shi et al. \shortcite{Shi20} implemented a library of differentiable filter nodes in the MATch system to enable a gradient-based optimization of existing procedural graphs to match a target appearance. MATch is however limited to differentiable operations, such as \emph{Filter} nodes and cannot optimize non-differentiable operations such as most \emph{Generator} nodes, which rely on discrete parameters with non-differentiable effects e.g.,  level of randomness of intensity/angles/sizes of patterns. This inherently prevents MATch from adjusting for structural differences between target materials and existing procedural graph output, limiting the generality of this solution, as shown in Figs. \ref{fig:newTeaser}\&\ref{fig:syn_results}.

To overcome this non-differentiability limitation, we propose leveraging deep learning to enable end-to-end gradient-based optimization of a procedural material. Our method allows joint optimization of all procedural graph parameters to match a material sample in terms of both structure and appearance.

The core idea is to use a \emph{Differentiable Proxy}, implemented as a differentiable neural network, to approximate a non-differentiable procedure. We adopt the state-of-the-art generative-model-like architecture StyleGAN2 and adapt its inputs and loss to our application. %
By replacing the original non-differentiable procedures with differentiable proxies, we create a differentiable space to optimize all parameters, including discrete ones. More generally, such proxies allow differentiation of black-box functions for which clearly defined gradients do not exist or cannot be determined explicitly.

We propose a three-stage strategy to smooth the optimization and avoid local minima. The pre-optimization step calibrates the material appearance for better structure initialization. The global optimization starts by efficiently finding a good generator initialization and then jointly optimizes all parameters in the graph using our differentiable Proxy. The post-optimization further refines the material by switching back to the original non-differentiable generator, using the optimal generator parameters found in the last step, and refining only differentiable parameters.

We experiment with different differentiable proxies with both regular and stochastic generation patterns. We show that our neural approximation well reproduces visual patterns created by the original generators. Integrating our neural proxy, we analyze our novel optimization routine and show that we can match the appearance of challenging materials without manual hand-tuning. %

In summary, we propose a more general inverse modeling framework for material graphs through the following contributions:
\begin{itemize}[leftmargin=15pt]
    \item Neural differentiable proxies for non-differentiable procedures.
    \item A multi-stage optimization strategy, achieving a high-quality match of a procedural material to a target appearance.
    \item An end-to-end fully differentiable pipeline that is more general than previous approaches, allowing optimization of all graph parameters without manual tuning.
\end{itemize}
The code is available at \url{https://github.com/yiwei-hu/DiffProxy}.
\section{Related Work}
\subsection{Material Procedural Modeling}

Procedural modeling of materials aims at representing analytic materials as procedures \cite{SubstanceDes, GuoBayesian20, Liu16, Guehl20, hu2019, hu2022, Shi20}. These methods generate procedural content from an image, materials or simply imagination. Closest to our method are works by  Hu et al.~\shortcite{hu2019, hu2022} and Shi et al.~\shortcite{Shi20} which estimate parameters of a given procedural graph to match an input material photo. 

Hu et al.~\shortcite{hu2019} train a neural network for each procedural materials, only learning to predict their artist-exposed parameters as opposed to all node graph parameters.

Shi et al.~\shortcite{Shi20} (MATch) implemented a differentiable version of filter nodes in the Substance Engine \shortcite{SubstanceDes} to optimize their continuous parameters to match a target material appearance. Being limited to filter nodes, MATch can only optimize the material appearance (e.g., albedo, roughness) and fails to match material structure.

Hu et al.~\shortcite{hu2022} present a semi-automatic pipeline for creating a material graph given material maps, requiring artists to segment the material they want to proceduralize. %
Hu et al.'s non-differentiable structure matching is disconnected from material property optimization, preventing joint optimization. Additionally, their structure matching step requires a time-consuming ($\sim$20 min) gradient-free optimization, dominating their runtime.

Our fully differentiable framework addresses the non-differentiability of generator nodes and can be directly plugged into MATch \cite{Shi20} and Hu et al.~\shortcite{hu2022} method, enabling end-to-end global optimization with better and faster appearance matching.

\subsection{Material Acquisition}
Material acquisition targets the recovery of material properties based on one or more images. Traditionally, dozens to thousands of images were required to sample the light-view space as described in the extensive survey by Guarnera et al.~\shortcite{Guarnera16}. More recently, deep neural networks were used to improve reconstruction from a single image \cite{Li17, Deschaintre18, henzler21neuralmaterial, Guo21, Zhou21} and from a small number of images \cite{Deschaintre19, Gao19, Guo20, Deschaintre20, Ye21}.  These methods can be separated into two categories. The first category relies on a single forward inference to recover the material parameters using an encoder/decoder architecture \cite{Li17, Deschaintre18, Deschaintre19, Zhou21, Guo21, Ye21}, while the second optimizes the latent space of a pre-trained decoder network  \cite{Gao19, Guo20, henzler21neuralmaterial}. While these methods can recover material parameters, they primarily focus on the reconstruction of material parameter pixel maps with their limited resolution and editability. In contrast to pixel maps, we focus on procedural representations of materials which allow users to parametrically control, edit and synthesize  materials at any resolution and scale. 

\subsection{Program Generation}
Since a material graph can be interpreted as a program, approaches that generate or infer programs from a given input are relevant to our work. Most of the research in program generation and inference has focused on the 2D or 3D shape domain, with a few notable exceptions~\cite{hu2018exposure, ganin2018synthesizing} which generate sequences of image edits using reinforcement learning. Early works focus on inverse proceduralization of given 3D shapes~\cite{demir2016proceduralization, vst2010inverse}, while more recent work uses data-driven methods to learn a prior over shape programs that can either be used for program generation or program induction from a given shape~\cite{du2018inversecsg, NEURIPS2019_50d2d226, NEURIPS2018_67880768, Sharma_2018_CVPR, tian2018learning, lu2019neurally, jones2020shapeAssembly, johnson2017inferring, walke2020learning, kania2020ucsgnet, xu2021inferring, wu2019carpentry}.

These methods operate in a different domain than our approach and also differ in their problem setup. Our goal is to modify an existing program, i.e., a computational graph, to more closely match a target material image rather than generating a program from scratch. 
We focus on inferring parameter configurations of a graph.

\section{Method}
Procedural material graphs are acyclic computational graphs generating spatially-varying (SV) material maps. The starting point of a material graph is a set of \textit{generators} producing patterns at multiple scales which serve as building blocks of a material graph. These generators typically define structures and multi-level texture features of the material, like a brick or tile pattern. Starting from the generator outputs, a variety of different \textit{filters} modify their appearance and combine them in multiple steps to achieve a final realistic procedural material. Although the generators play an important role in material graph design, previous graph optimization frameworks assume that the generators are pre-calibrated and fixed during graph optimization \cite{Shi20, hu2022}, thus they cannot optimize some features of the material. In contrast, we propose a fully end-to-end optimization framework that allows global and joint optimization of both material structures/features and material appearance. 

The main challenge for optimizing generator nodes is their non-differentiablity. Our key idea is to avoid direct optimization of the original generator by optimizing a differentiable proxy.

\begin{figure}
	\centering
	\addtolength{\tabcolsep}{-4pt}
	\begin{tabular}{ccc}
		\includegraphics[width=0.155\textwidth]{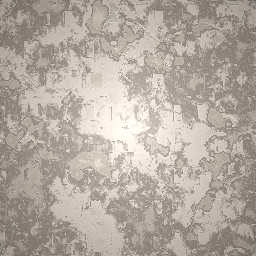}\llap{\frame{\includegraphics[width=0.07\textwidth]{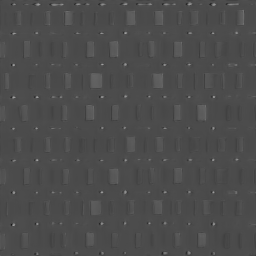}}} &
		\includegraphics[width=0.155\textwidth]{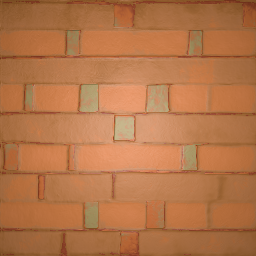}\llap{\frame{\includegraphics[width=0.07\textwidth]{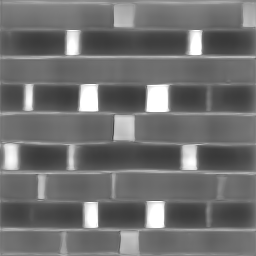}}} &
		\includegraphics[width=0.155\textwidth]{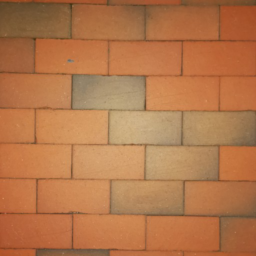} \\
		Init. & StyleGAN2 Optim. & Target \\
        \includegraphics[width=0.155\textwidth]{figures/image/bad_proxies/init.png}\llap{\frame{\includegraphics[width=0.07\textwidth]{figures/image/bad_proxies/init_mask.png}}} &
		\includegraphics[width=0.155\textwidth]{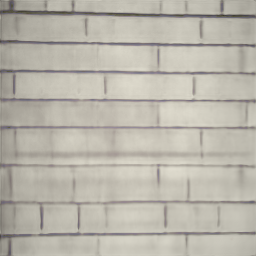}\llap{\frame{\includegraphics[width=0.07\textwidth]{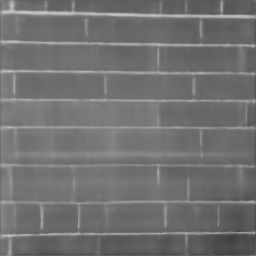}}} &
		\includegraphics[width=0.155\textwidth]{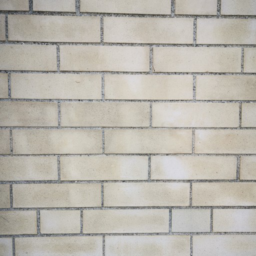} \\
	\end{tabular}
\caption{Result using the original StyleGAN2 architecture as differentiable proxy. Insets represent generator maps synthesized by the trained proxy before/after optimization. We follow the procedure described in Sec. \ref{Sec:optimization} to optimize the structure using StyleGAN2's $W+$ latent space. We see that it fails to generate a good pattern to match the target. Our proxy solves these issues as can be seen in Fig.~\ref{fig:newTeaser} and supplemental materials.}%
\label{fig:bad_proxies}
\Description{Bad Proxies (StyleGAN2)}
\end{figure}

\begin{figure}[t]
	\centering
	\addtolength{\tabcolsep}{-4pt}
	\begin{tabular}{cccc}
		\includegraphics[width=0.155\textwidth]{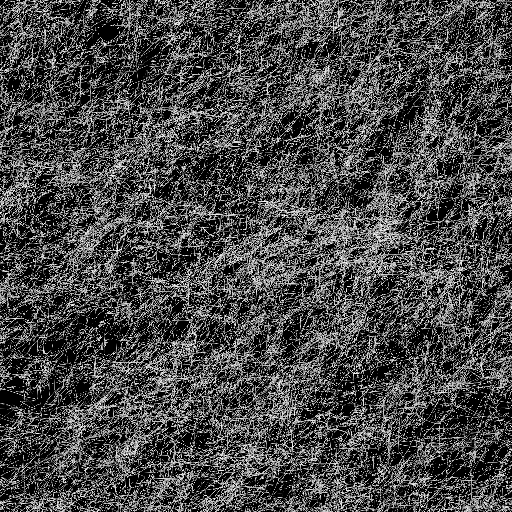} &
		\includegraphics[width=0.155\textwidth]{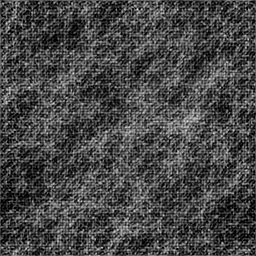} &
		\includegraphics[width=0.155\textwidth]{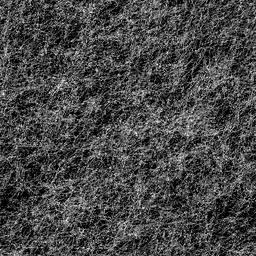} \\
		\includegraphics[width=0.155\textwidth]{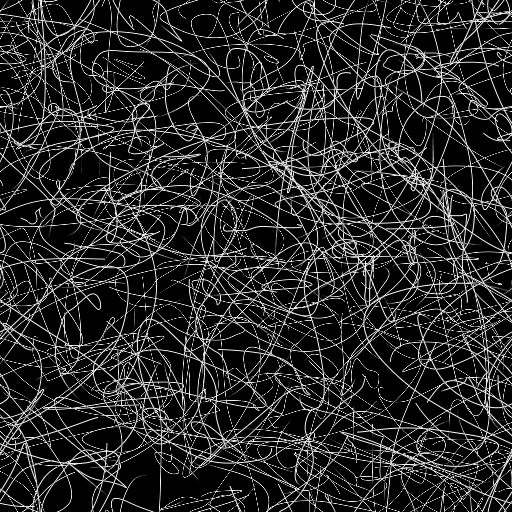} &
		\includegraphics[width=0.155\textwidth]{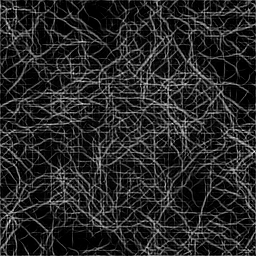} &
		\includegraphics[width=0.155\textwidth]{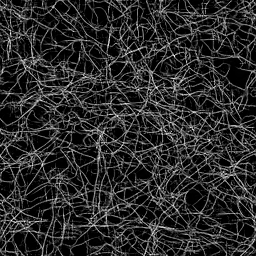} \\
		Real & w/o adv. loss & w/ adv. loss \\
	\end{tabular}
\caption{Adding an adversarial loss helps reduce artifacts when patterns are stochastic. We show here a typical case with the scratches generator which generates highly stochastic patterns.}
\label{fig:gan_loss}
\Description{With vs without GAN loss}
\end{figure}
\begin{figure} %
	\centering
	\addtolength{\tabcolsep}{-3pt}
	\begin{tabular}{ccccc}%
		& Real & Approx. & Real & Approx.\\
		\raisebox{25pt}{\scalebox{0.8}{\rotatebox[origin=c]{90}{Brick}}} &
		\includegraphics[width=0.2225\columnwidth]{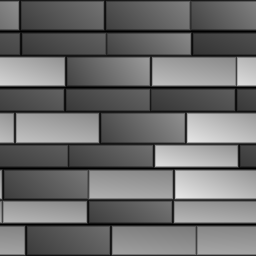} & 
        \includegraphics[width=0.2225\columnwidth]{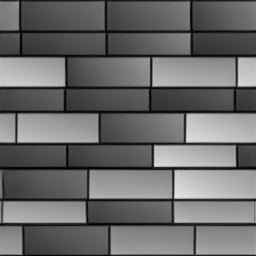} &
		\includegraphics[width=0.2225\columnwidth]{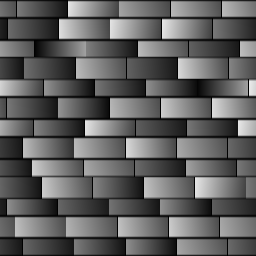} & 
        \includegraphics[width=0.2225\columnwidth]{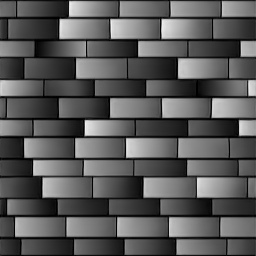} \\
		\raisebox{25pt}{\scalebox{0.8}{\rotatebox[origin=c]{90}{Scratch}}} &
		\includegraphics[width=0.2225\columnwidth]{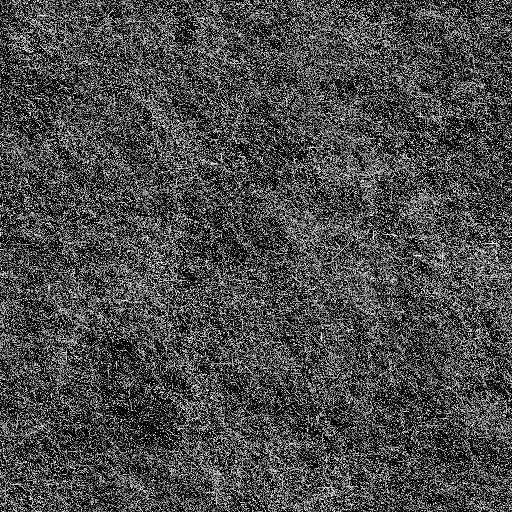} & 
        \includegraphics[width=0.2225\columnwidth]{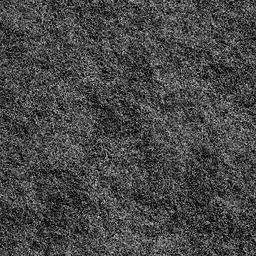} &
		\includegraphics[width=0.2225\columnwidth]{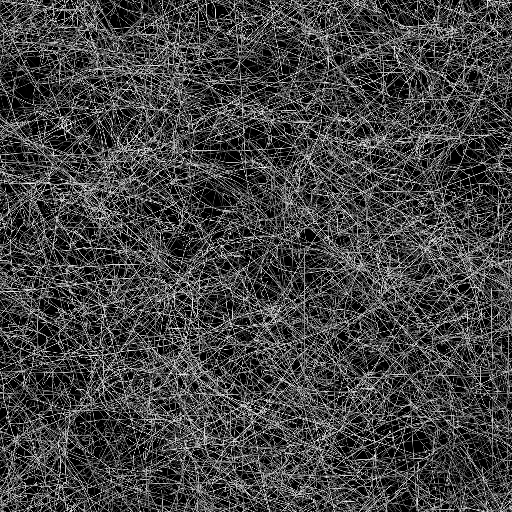} & 
        \includegraphics[width=0.2225\columnwidth]{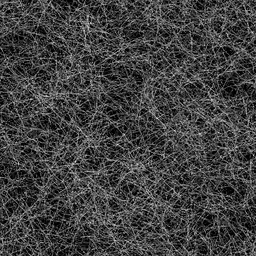} \\
		\raisebox{25pt}{\scalebox{0.8}{\rotatebox[origin=c]{90}{Tile (Paraboloid)}}} &
		\includegraphics[width=0.2225\columnwidth]{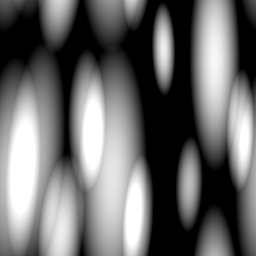} & 
        \includegraphics[width=0.2225\columnwidth]{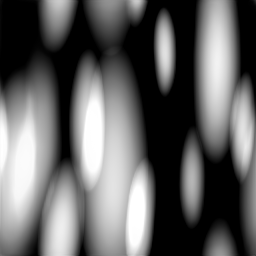} &
		\includegraphics[width=0.2225\columnwidth]{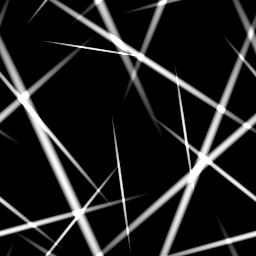} & 
        \includegraphics[width=0.2225\columnwidth]{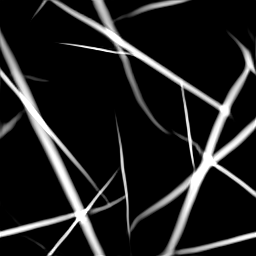} \\
		\raisebox{25pt}{\scalebox{0.8}{\rotatebox[origin=c]{90}{Arc Pavement}}} &
		\includegraphics[width=0.2225\columnwidth]{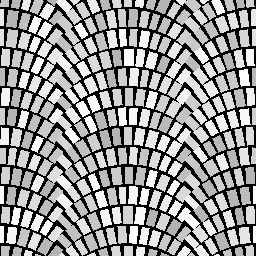} & 
        \includegraphics[width=0.2225\columnwidth]{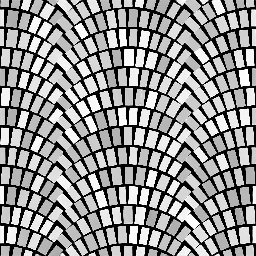} &
		\includegraphics[width=0.2225\columnwidth]{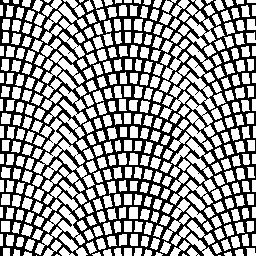} & 
        \includegraphics[width=0.2225\columnwidth]{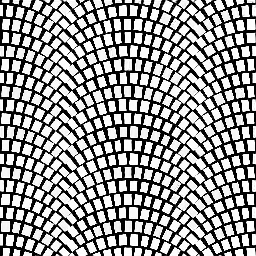} \\
		\raisebox{25pt}{\scalebox{0.8}{\rotatebox[origin=c]{90}{PPTBF}}} &
		\includegraphics[width=0.2225\columnwidth]{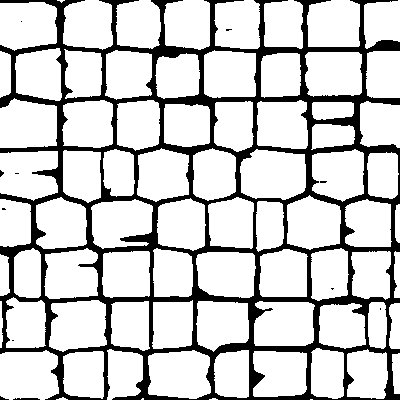} & 
        \includegraphics[width=0.2225\columnwidth]{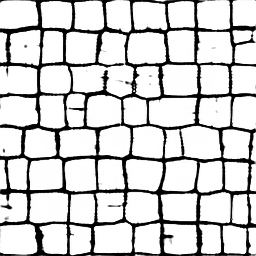} &
		\includegraphics[width=0.2225\columnwidth]{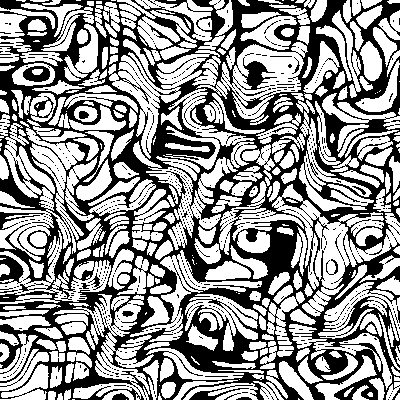} & 
        \includegraphics[width=0.2225\columnwidth]{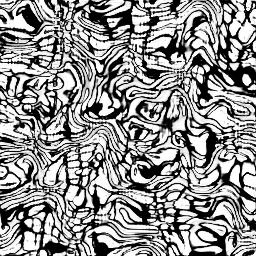} \\
	\end{tabular}
\caption{We compare generator maps synthesized by our proxies (Approx.) with their original procedural counterpart (Real) via randomly sampled parameters, showing that they are very close. See supplemental documents for results generated from all our trained proxies.} %
\label{fig:proxies}
\Description{Results of the differentiable proxies}
\end{figure}
\subsection{Differentiable Proxy} \label{Sec:proxy}
Given an arbitrary, non-differentiable, 2D image generator $G(\theta)$, where $\theta$ represents its procedural parameters, we create a differentiable proxy. To do so, we approximate $G(\theta)$ with a convolutional neural network (CNN) $\hat{G}(\theta)$ trained to reproduce the behaviour of $G(\theta)$: given a set of parameters $\theta_i$ we want $G(\theta_i) = \hat{G}(\theta_i)$.

A natural solution would be to train a generative model (e.g., StyleGAN2). However, we notice that generative models' optimizable latent spaces (e.g., $W+$ in StyleGAN2) can be expressive beyond the original generator scope. This allows the proxy to generate maps that do not exist in the original generator space, leading to (1) poor appearance matching and (2) complicating the process of mapping back to the original generator parameters. The problem (1) is illustrated in  Fig.~\ref{fig:bad_proxies}. We trained the original StyleGAN2 with the data sampled using the process described in Sec.~\ref{sec:implementation}. The insets of the second column of Fig.~\ref{fig:bad_proxies} represent the direct outputs of the trained StyleGAN2 which fail to match the targets. Solving problem (2) in this context would require training an additional network mapping from $W+$ to the original generator parameter space.  A more detailed discussion of the limits of using StyleGAN2 or AutoEncoder~\cite{Gao19} is available in the supplemental materials.

To overcome this problem, we use a modified StyleGAN2~\cite{stylegan2} architecture. We modify the architecture input and train our network to approximate the mapping between the parameters and the output of a generator node.
Specifically, rather than starting from a random latent vector $Z$, we encode parameters $\theta$ into an intermediate latent space $W$, through a set of fully connected layers. We then feed $W$ to AdaIn layers, similar to the original StyleGAN.

We further modify the architecture to use $\theta$ as its sole input: we remove the noise added in each block of the original StyleGAN2, making our network deterministic, allowing it to learn the one-to-one mapping as we need.

We train our neural proxy using synthetic data, directly generated by carefully sampling (Sec.~\ref{sec:implementation}) the procedural generator.
With our modified architecture, we can train our differentiable proxy $\hat{G}$ by directly sampling a data pair $(\theta, I)$ and minimizing the per-pixel difference between the approximated generator map $\hat{G}(\theta)$ and the ground-truth $I$. We design a weighted loss function combining $L_1$ loss, deep feature loss, style loss~\cite{gatys2015}, and an optional adversarial loss $L_{\text{Adv}}$:
\begin{equation}
    L = \lambda_0 L_1 + \lambda_1 L_{\text{feat}} + \lambda_2 L_{\text{style}} ( + \lambda_3 L_{\text{Adv}})
\end{equation}
Deep feature loss $L_{\text{feat}}$ is defined by the $L_1$ difference between deep feature maps extracted from a pre-trained VGG19 (\cite{vgg19} neural network. Style loss $L_{\text{style}}$ is defined by the $L_1$ difference between the Gram Matrices of extracted deep feature maps. We empirically assign $\lambda_0 = \lambda_2 = 1$ and $\lambda_1=10$.

In our experiments, the first three loss terms are generally enough to guide the neural network to learn a one-to-one mapping for procedural generators ($\lambda_3 = 0$). However, for highly random patterns, such as scratch generators, this combination of losses results in small artifacts and struggles to reproduce stochastic behavior. To solve this, we add an adversarial critic to improve the fitting quality, as shown in Fig. \ref{fig:gan_loss}. For these stochastic generators, we adjust $\lambda_3 > 0$ to ensure that the loss is still dominated by the first three loss terms, relaxing the one-on-one mapping goal, without replacing it. $L_{\text{Adv}}$ is defined as the cross-entropy loss with $R1$ regularization \cite{Mescheder2018}. The critic takes both parameter $\theta$ and generator map $I$ as input and evaluates whether it is a real or fake data pair. 

\begin{figure}
	\centering
	\addtolength{\tabcolsep}{-4pt}
	\begin{tabular}{cccc}
		\includegraphics[width=0.155\textwidth]{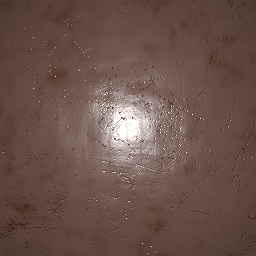} &
		\includegraphics[width=0.155\textwidth]{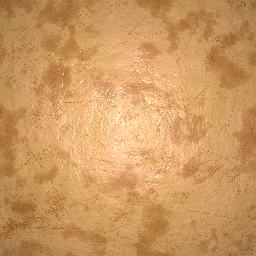} &
		\includegraphics[width=0.155\textwidth]{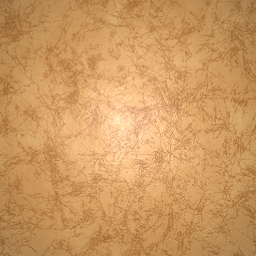} \\
		Input & MATch (Stage I) & Stage II \\
		\includegraphics[width=0.155\textwidth]{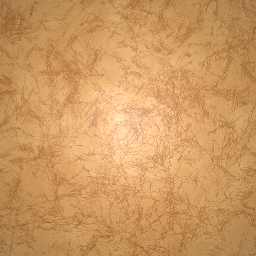} &
		\includegraphics[width=0.155\textwidth]{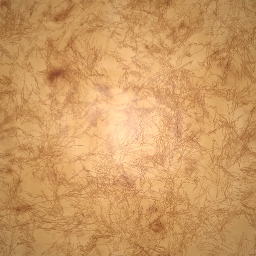} &
		\includegraphics[width=0.155\textwidth]{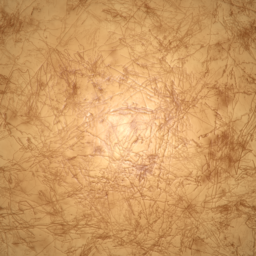} & \\
         Stage II* & Refined & Target \\
	\end{tabular}
\caption{We optimize a leather material (Input) to match a scratched potato skin (Target). We first match the overall material parameters such as color or roughness (MATch, Stage I). After global optimization (Stage II), we retrieve correct scratch patterns. We then replace our proxy with the real generator (Stage II*) and re-optimize the filter nodes with fixed generators and a smaller learning rate, refining the result (Refined) to best match the target.}
\label{fig:refinement}
\Description{Illustration of the post-refinement step}
\end{figure}
\subsection{Our Differentiable Proxies}
In our experiments, we train differentiable proxies for generators in both MATch and the more recent system by Hu et al. \shortcite{hu2022}. In the case of generators for MATch we train proxies for the following generators, which represent the majority of generators used across 100 analyzed Substances: \emph{Brick Generator, Stripe Generator, Scratches Generator, Tile Generator (Paraboloid), Tile Generator (Brick), Arc Pavement Generator}. 
In the case of Hu et al.~\shortcite{hu2022} we  train a proxy for the generator they used: \emph{Point Process Texture Basis Functions (PPTBF)}~\cite{Guehl20}).

We compare in Fig.~\ref{fig:proxies} generator maps synthesized with our trained differentiable proxy $\hat{G}$ to those generated with the original procedural generator $G$. Each result shows a randomly sampled set of parameters, showing that our proxies can generate mask maps which closely approximate those generated by $G$. This ensures that we remain in the space of masks that can be generated by $G$ during proxy optimization, allowing us to project back the parameters to the original non-differentiable procedural material graph.

\subsection{Fully Differentiable Optimization} \label{Sec:optimization}
Given a user (or classifier~\cite{Shi20}) chosen material graph $\mathcal{G}$, we convert it to its differentiable counterpart by replacing the procedural generator nodes with our %
proxies and the filters with differentiable filter nodes from the DiffMat library~\cite{Shi20}. The optimizable parameters are $\theta=(\theta_g, \theta_f)$ where $\theta_g$ and $\theta_f$ drive our differentiable proxies and the differentiable filters respectively.

The material graph $\mathcal{G}$ outputs a set of spatially-varying 2D material maps $M=\mathcal{G}(\theta)$ (e.g., albedo, normal, roughness, and metallic). A rendering operator $R$ can be applied to synthesize an image $I=R(M)$. Our optimization process recovers an optimal $\theta^*$ that minimizes the difference $d(I, I^*)$ between our rendered image $I=R(M)$ and a user provided target image $I^*$:
\begin{equation}
    \theta^*=\operatorname*{argmin}_\theta L_{\theta} = \operatorname*{argmin}_\theta d(I, I^*)
\end{equation}
Leveraging $\mathcal{G}$ differentiability, we optimize $\theta^*$ with gradient descent using PyTorch \cite{pytorch} as a general auto-differentiation framework. 

A key challenge is the existence of local minima
in the joint optimization of both generators and filters. While we relax the generator differentiability, the original discrete attribute variation tends to form local minima, making the optimization non-convex. To stabilize it, we propose a multi-stage optimization strategy. 

\subsubsection{Stage I: Pre-optimization} We start by only matching the overall material appearance, i.e., only optimizing $\theta_f$ with a fixed $\theta_g$. This is a pre-optimization step. 
We use the MATch framework.
The loss function we defined is a multi-scale style loss:
\begin{equation}
    L_{\theta_f} = ||GM(I) - GM(I^*)||_1
\label{Eq:pre-optimization}
\end{equation} 
where $GM$ is an operator that computes Gram Matrices of extracted deep features \cite{gatys2015}. We compute and aggregate the loss function at multiple resolutions (256x256, 128x128 and 64x64). This pre-optimization step calibrates basic material properties (e.g., albedos and roughness). In few cases MATch fails to improve the initialization, in which case, we directly move to Stage II.

\subsubsection{Stage II: Global Optimization} We use the optimization results from Stage I to initialize this step. We now optimize the entire set of parameters $\theta = (\theta_f, \theta_g)$. To minimize the impact of local minima, we find a good initialization. We randomly sample possible generator parameters and initialize our optimization with the parameters which generate the closest appearance to the targeted image:
\begin{equation}
    L_{\theta_g} = ||F(I_g) - F(I_g^*)||_1
\end{equation}
where $F$ denotes the extracted deep features from a pre-trained neural network \cite{vgg19}. $I_g$ and $I_g^*$ are grayscale version of $I$ and $I^*$ respectively. We can efficiently compare the grayscale images as color and roughness values were already optimized. In practice, we sample 500 possible parameters initialization for a differentiable graph $\mathcal{G}$ in less than a minute.

Using the selected initialization of $\theta_g$, we optimize all parameters $\theta$ with a combination of feature and style loss:
\begin{equation}
    L_{\theta} = ||F(I) - F(I^*)||_1 + \alpha ||GM(I) - GM(I^*)||_1
\end{equation}
where $\alpha$ is a weighting variable. We use feature loss as a main loss term to measure the structure and appearance similarity. We also add a small style loss component as a flexible component that matches the overall statistics of material appearance as the procedural material may not always achieve pixel-perfect matches of real material pictures. We empirically choose $\alpha=0.05$ and, similar to Eq.~\ref{Eq:pre-optimization}, we evaluate the loss function using a multi-resolution approach. To further reduce the risk of encountering local minima, we set a larger learning rate, decaying over the optimization.

\subsubsection{Stage III: Post-Optimization} As our differentiable proxies may generate slightly different images than the original generators, we fine-tune the results of our Global Optimization step.
In this step, we switch our optimized proxies back to their original procedural counterparts with the optimized parameters $\theta_g$. As the generators are now non-differentiable, we fine-tune the filter nodes parameters $\theta_f$. %
This fine-tuning step is the same as Stage I, using a multi-scale style loss as described in Eq. \ref{Eq:pre-optimization}, with a closer material structure and a finer learning rate, refining the appearance to better match the target. We show the result of each of these step in Fig.~\ref{fig:refinement}, and videos of the optimization process in the supplemental material.

\begin{figure}
	\centering
	\addtolength{\tabcolsep}{-4pt}
	\begin{tabular}{ccccc}
		\includegraphics[width=0.115\textwidth]{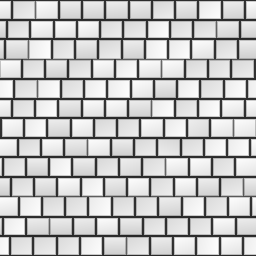} &
		\includegraphics[width=0.115\textwidth]{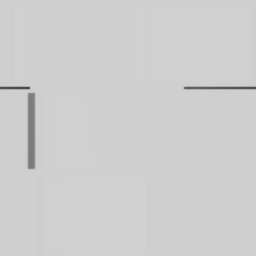} &
		\includegraphics[width=0.115\textwidth]{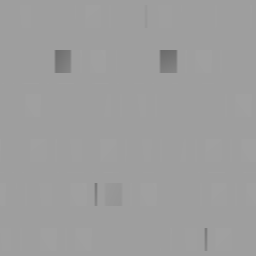} &
		\includegraphics[width=0.115\textwidth]{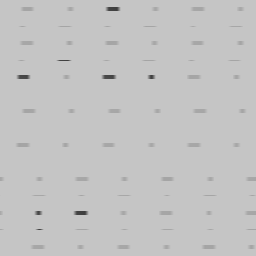} \\
		Artists' Space & \multicolumn{3}{c}{General Generator Space}
	\end{tabular}
\caption{We sample the data to train our proxies by sampling in the parameter spaces used by artists. On the left is an example of pattern generated from this space, while on the right are three examples of (undesirable) patterns that can be generated with the same generator, with uniform sampling of the entire parameter space.}
\label{fig:bad_map}
\Description{Importance-sampling of parameters}
\end{figure}
\section{Implementation}
\label{sec:implementation}
To train our models, we build a dataset for a chosen generator by randomly sampling its parameter space. However, a fully random-sampling strategy leads to unrealistic or invalid patterns. For instance, some parameter combinations for a brick pattern generator may not lead to a brick-like pattern, as shown in Fig.~\ref{fig:bad_map}.

To build a more representative dataset, we use a heuristic sampling approach. For popular systems, we leverage existing collections of node graphs. For example when training proxies compatible with MATch, (based on Substance), 
we analyze the parameter distribution from Substance Source, a database containing 7000+ artist-designed material graphs~\cite{SubstanceDes}. We compute the range of parameters used by artists, as well as their mean and standard derivations, and independently sample each parameter. %
We generate a 300,000 256x256 images dataset for each differentiable proxy $\hat{G}$, which, depending on the procedural generator complexity takes $5\sim10$ hours. Once trained, each $\hat{G}$ requires less than 100 MB.

To train the PPTBF proxy for the recent Hu et al.~\shortcite{hu2022} framework, we use the PPTBF released dataset \cite{Guehl20} and resample the parameters based on the parameter distributions analyzed from that dataset. We sample an additional 500,000 mask maps to increase the sample density.

We implement our differentiable proxies and optimization pipeline in PyTorch. We train our proxy using a Nvidia RTX 3090 with a CPU of Intel i9 10900K. To reach convergence, 60 epochs of training usually takes $2\sim3$ days on a single GPU, using Adam \cite{Adam} with a learning rate of $0.0025$, a batch size of 32 and normalizing the parameters between $0$ and $1$ to stabilize the training. When applying adversarial training, we set $\gamma=10$ for $R1$ and train the discriminator with Adam and a learning rate of $10^{-3}$.%

For the parameter optimization itself, we use the PyTorch auto-differentiation framework %
and use Adam($\beta=(0.9, 0.999)$ with a learning rate decaying strategy (lr is halved every 200 steps) to optimize for our target appearance.

We set a smaller learning rate for pre- and post- optimization steps (0.002) because the optimization space of continuous parameters of filter nodes are %
more convex. We set a larger initial learning rate for global optimization (0.02) as we observe that a larger early step size helps avoid local minima. The full optimization takes around $3\sim5$ minutes for 1000 steps, depending on the complexity (numbers of nodes and connections) of the material graphs. For reference, MATch optimization takes $2\sim3$ min. 

\begin{figure*}
	\centering
	\addtolength{\tabcolsep}{-4pt}
	\begin{tabular}{ccccccccc}
	    Init (Proc.) & MATch (Proc.) & \textbf{Ours (Proc.)} & Target & & Init (Proc.) & MATch (Proc.) & \textbf{Ours (Proc.)} & Target \\
		\includegraphics[width=0.12\textwidth]{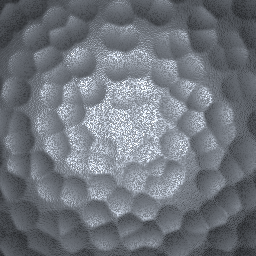} &
		\includegraphics[width=0.12\textwidth]{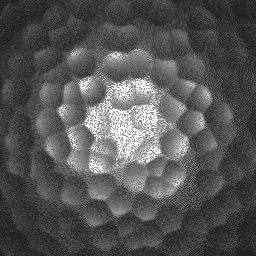} &
		\includegraphics[width=0.12\textwidth]{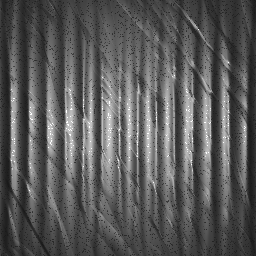} &
		\includegraphics[width=0.12\textwidth]{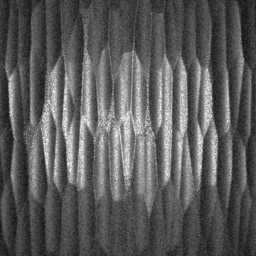} & &
		\includegraphics[width=0.12\textwidth]{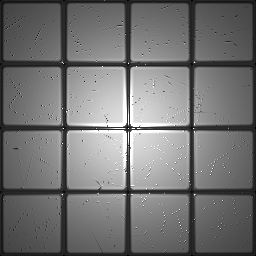} &
		\includegraphics[width=0.12\textwidth]{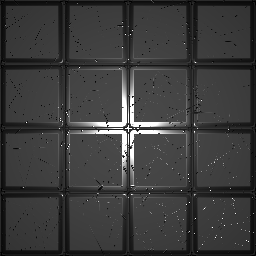} &
		\includegraphics[width=0.12\textwidth]{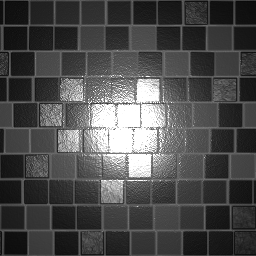} &
		\includegraphics[width=0.12\textwidth]{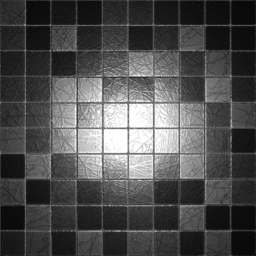} \\
		\includegraphics[width=0.12\textwidth]{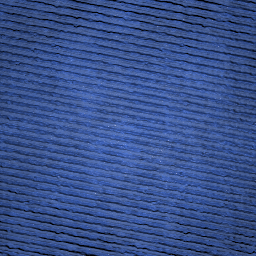} &
		\includegraphics[width=0.12\textwidth]{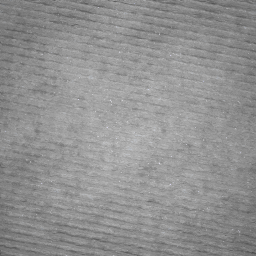} &
		\includegraphics[width=0.12\textwidth]{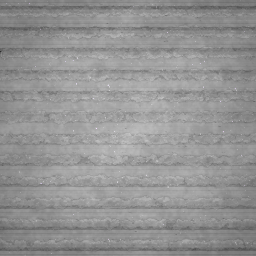} &
		\includegraphics[width=0.12\textwidth]{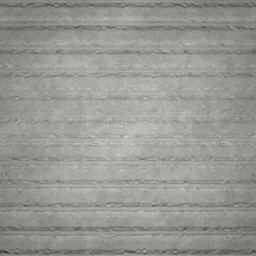} & &
		\includegraphics[width=0.12\textwidth]{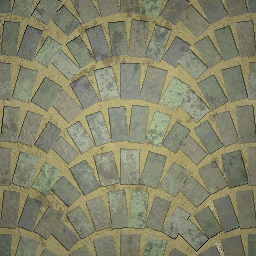} &
		\includegraphics[width=0.12\textwidth]{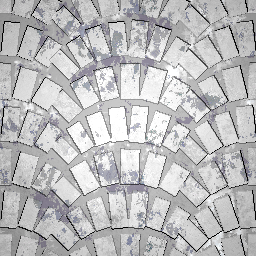} &
		\includegraphics[width=0.12\textwidth]{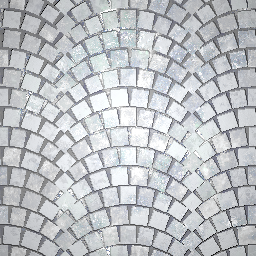} &
		\includegraphics[width=0.12\textwidth]{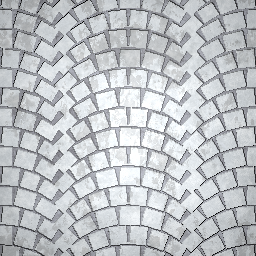} \\
		\includegraphics[width=0.12\textwidth]{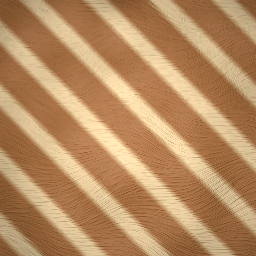} &
		\includegraphics[width=0.12\textwidth]{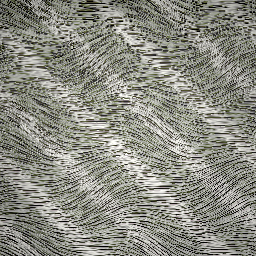} &
		\includegraphics[width=0.12\textwidth]{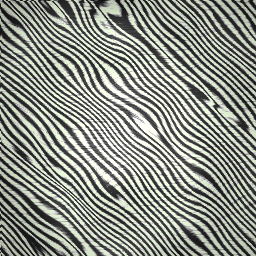} &
		\includegraphics[width=0.12\textwidth]{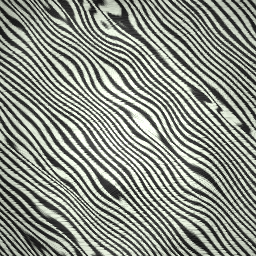} & &
		\includegraphics[width=0.12\textwidth]{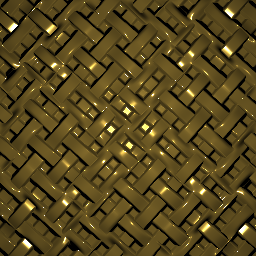} &
		\includegraphics[width=0.12\textwidth]{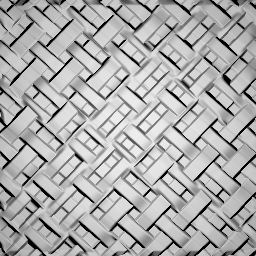} &
		\includegraphics[width=0.12\textwidth]{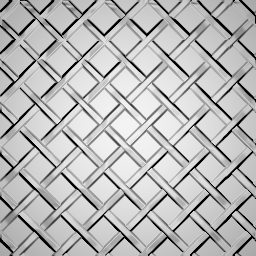} &
		\includegraphics[width=0.12\textwidth]{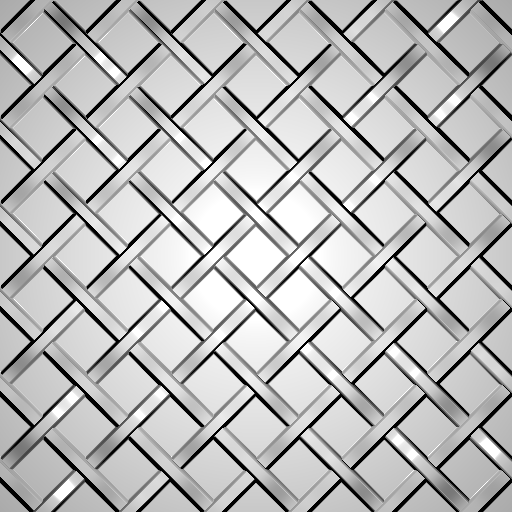}
	\end{tabular}
\caption{Results on synthetic materials. We sample random parameters in a graph as the target appearance and show that our optimization is able to recover these parameters even with a distant initialization. See supplemental materials for additional results.}
\label{fig:syn_results}
\Description{Optimization results on synthetic materials}
\end{figure*}
\begin{figure}
	\centering
	\addtolength{\tabcolsep}{-4pt}
	\begin{tabular}{ccc}
	        Input & Hu et al. & Ours \\
		    \includegraphics[width=0.155\textwidth]{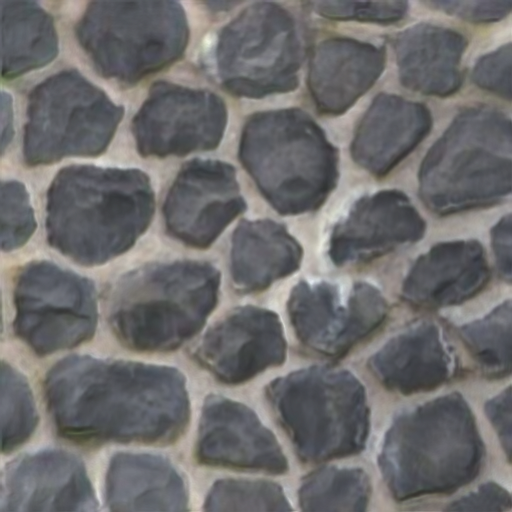}\llap{\frame{\includegraphics[width=0.07\textwidth]{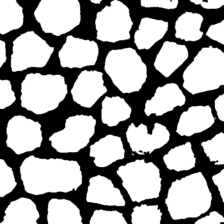}}} &
		    \includegraphics[width=0.155\textwidth]{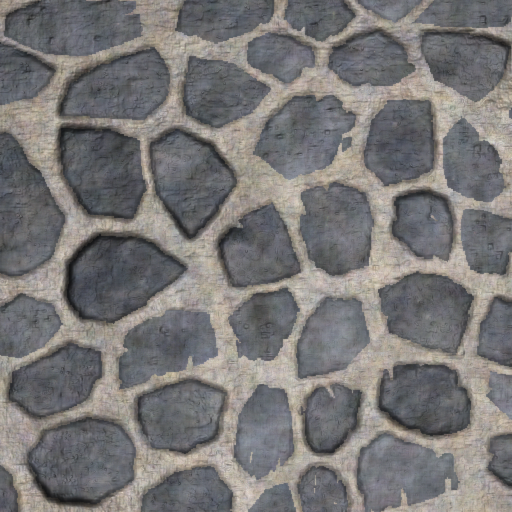}\llap{\frame{\includegraphics[width=0.07\textwidth]{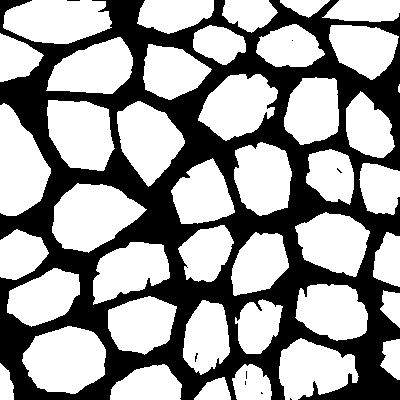}}} &
		    \includegraphics[width=0.155\textwidth]{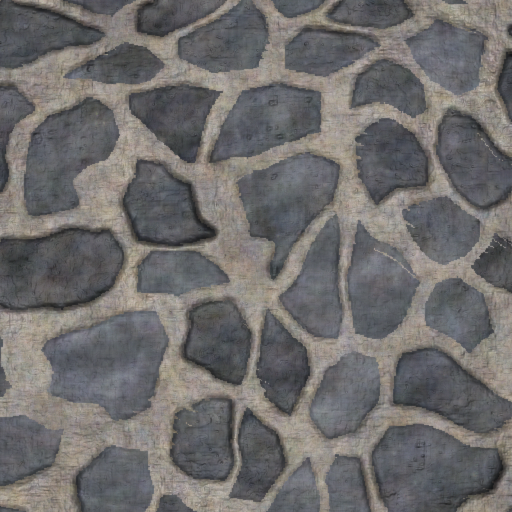}\llap{\frame{\includegraphics[width=0.07\textwidth]{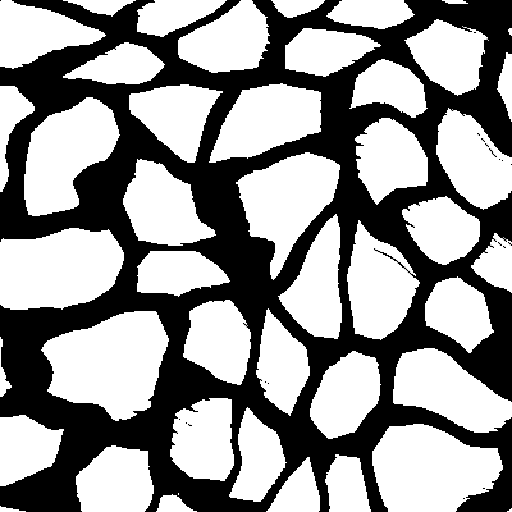}}}  \\
		    
		    \includegraphics[width=0.155\textwidth]{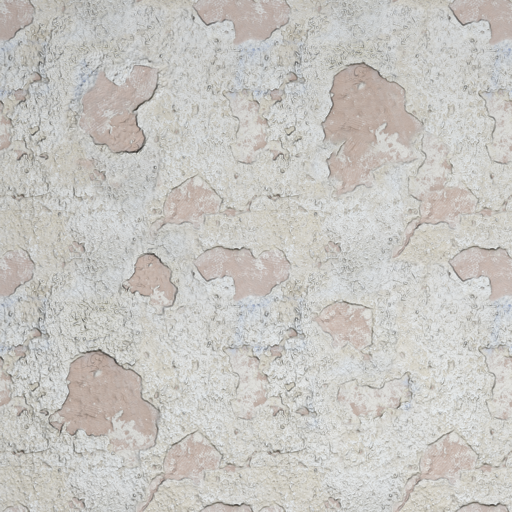}\llap{\frame{\includegraphics[width=0.07\textwidth]{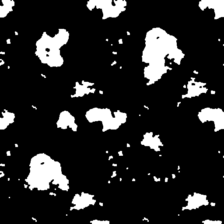}}} &
		    \includegraphics[width=0.155\textwidth]{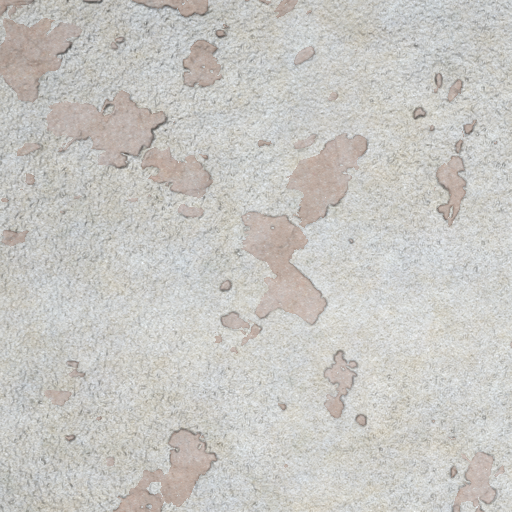}\llap{\frame{\includegraphics[width=0.07\textwidth]{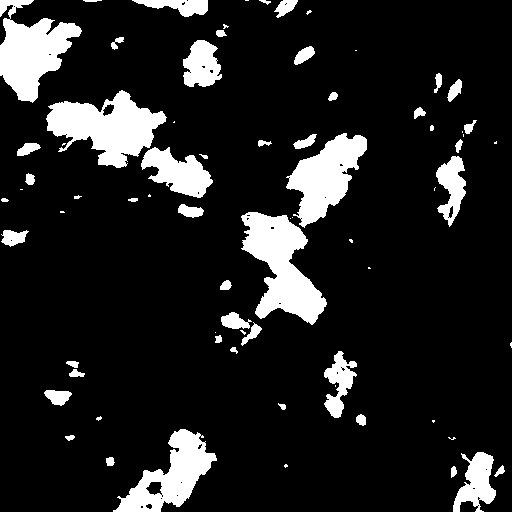}}} &
		    \includegraphics[width=0.155\textwidth]{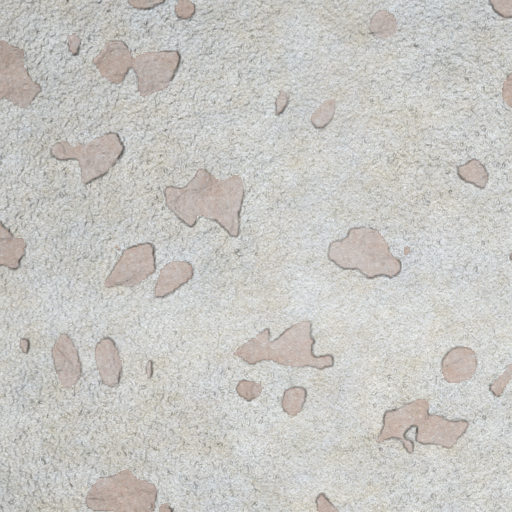}\llap{\frame{\includegraphics[width=0.07\textwidth]{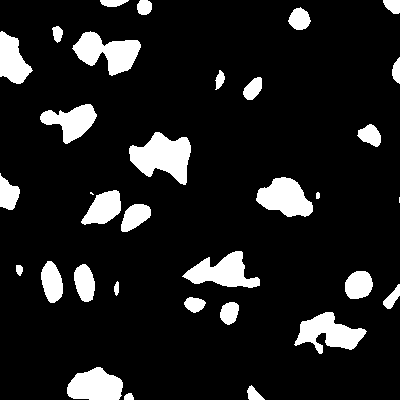}}}  \\
		    
		    \includegraphics[width=0.155\textwidth]{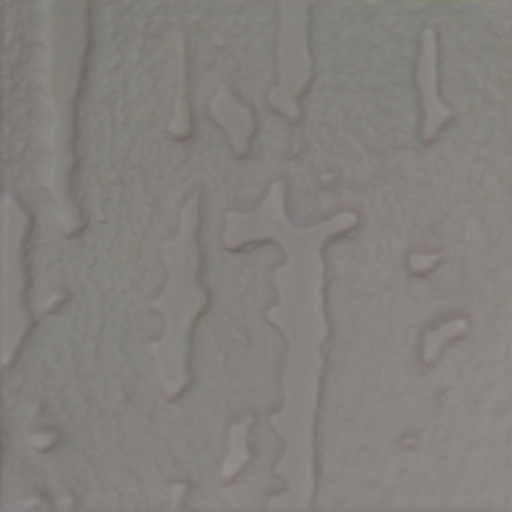}\llap{\frame{\includegraphics[width=0.07\textwidth]{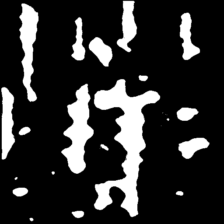}}} &
		    \includegraphics[width=0.155\textwidth]{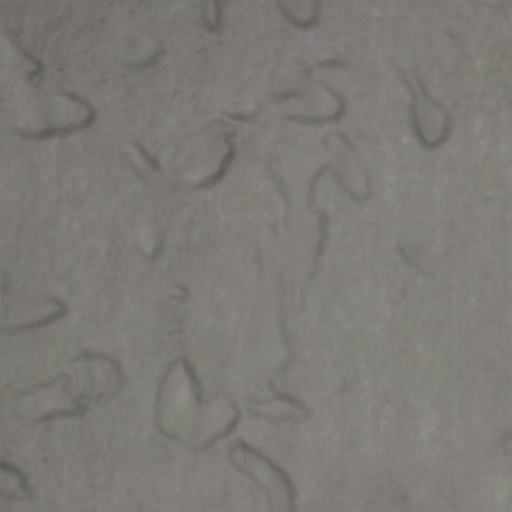}\llap{\frame{\includegraphics[width=0.07\textwidth]{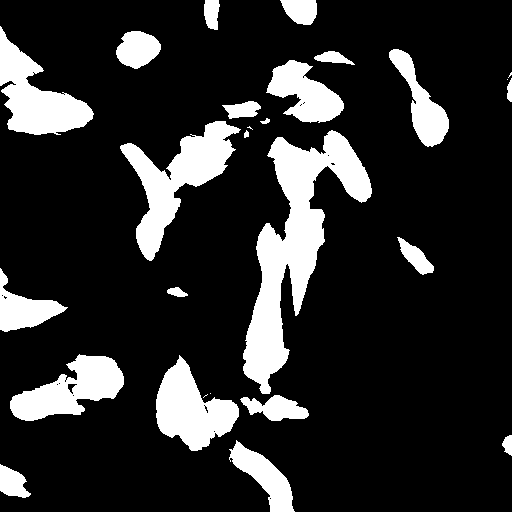}}} &
		    \includegraphics[width=0.155\textwidth]{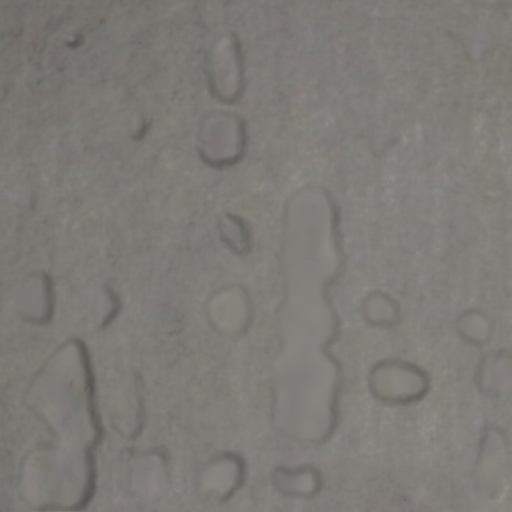}\llap{\frame{\includegraphics[width=0.07\textwidth]{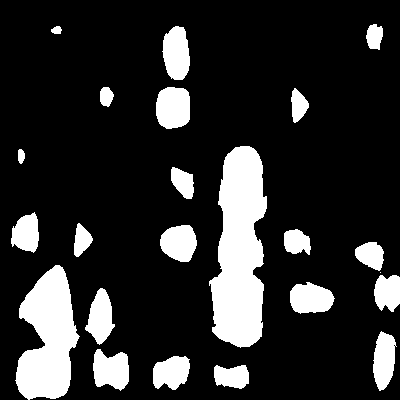}}}  \\
	\end{tabular}
\caption{We plug our differentiable PPTBF proxy in Hu et al.'s~\shortcite{hu2022} framework. Their framework optimizes parameters of a procedural PPTBF mask to match a user-segmented mask map with a gradient-free method, taking 20 minutes. Using SGD, enabled by our proxy, we achieve similar results in 30 seconds. See supplemental materials for additional comparisons.}
\label{fig:pptbf}
\Description{Comparisons to Hu et al. 2022}
\end{figure}

\begin{figure}
	\centering
	\addtolength{\tabcolsep}{-4pt}
	\begin{tabular}{ccccc}
		\includegraphics[width=0.115\textwidth]{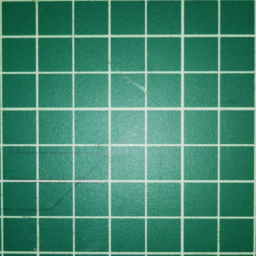} &
		\includegraphics[width=0.115\textwidth]{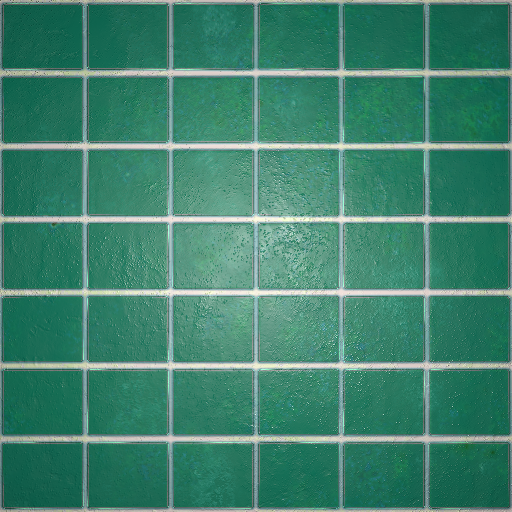} &
		\includegraphics[width=0.115\textwidth]{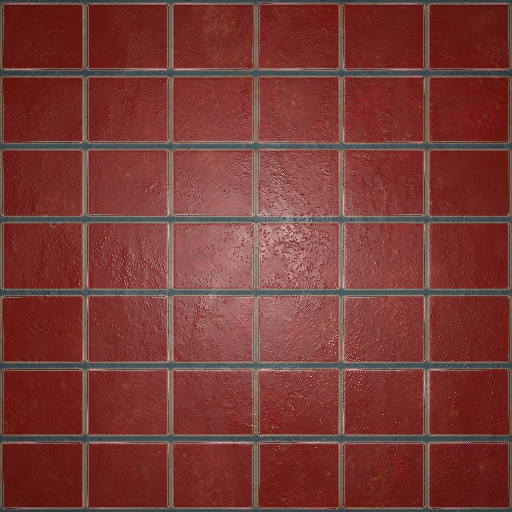} &
		\includegraphics[width=0.115\textwidth]{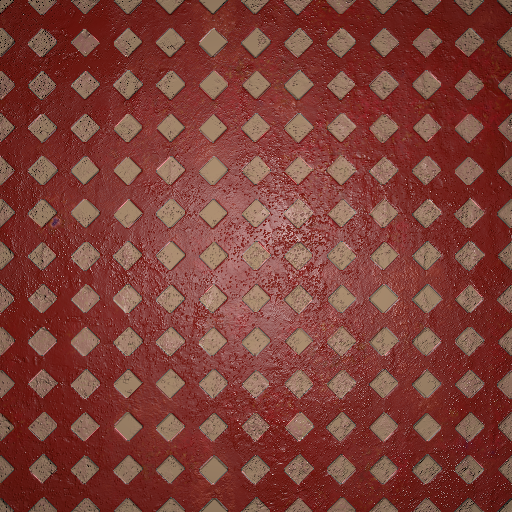} \\
		Target & Our Proc. & \multicolumn{2}{c}{Edits} \\
		\multicolumn{4}{c}{
		\includegraphics[width=0.47\textwidth]{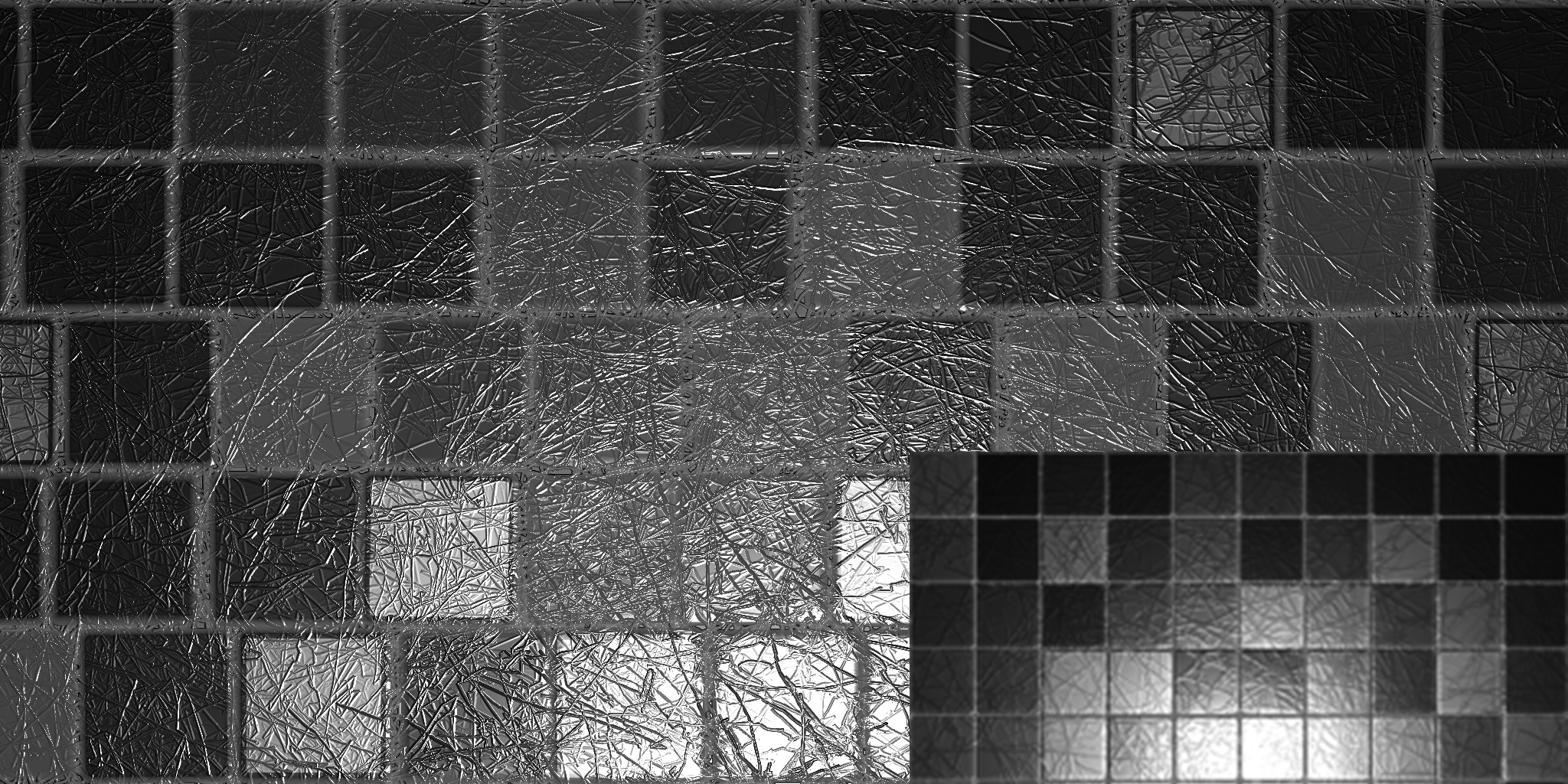}
		} \\
		\multicolumn{4}{c}{
		\includegraphics[width=0.47\textwidth]{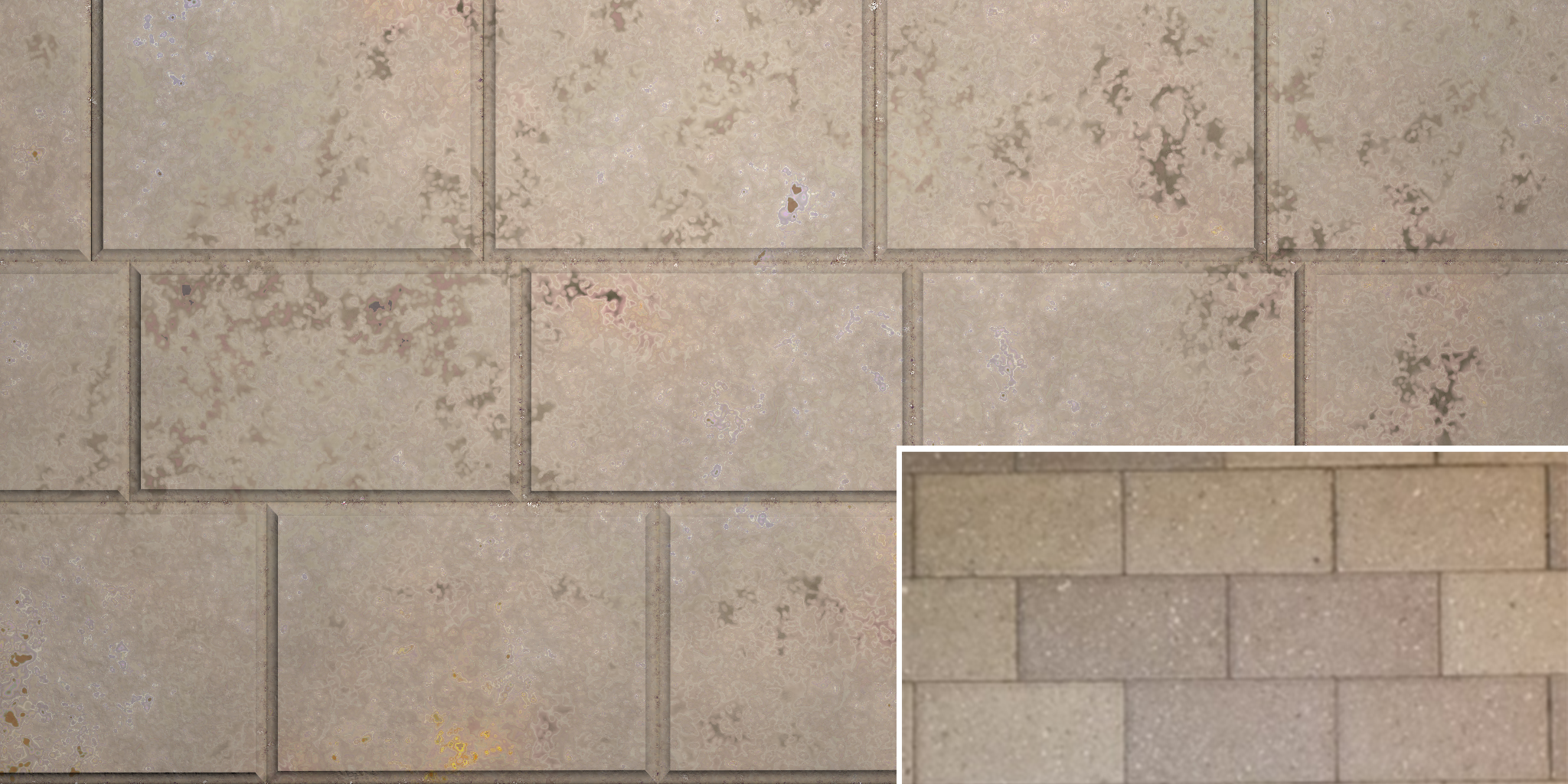}
		} \\
		\multicolumn{4}{c}{High Resolution Material} \\
		\multicolumn{2}{c}{
		\includegraphics[width=0.23\textwidth]{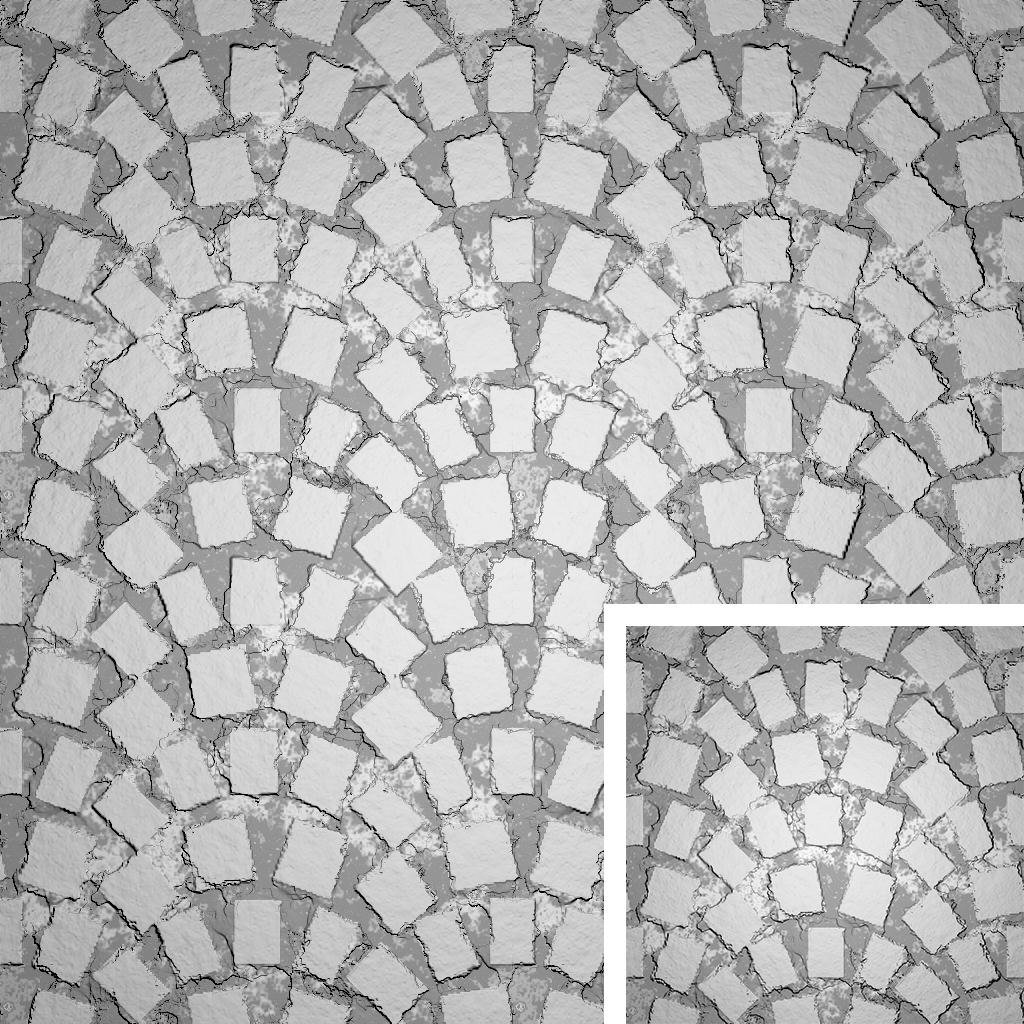}
		} &
		\multicolumn{2}{c}{
		\includegraphics[width=0.23\textwidth]{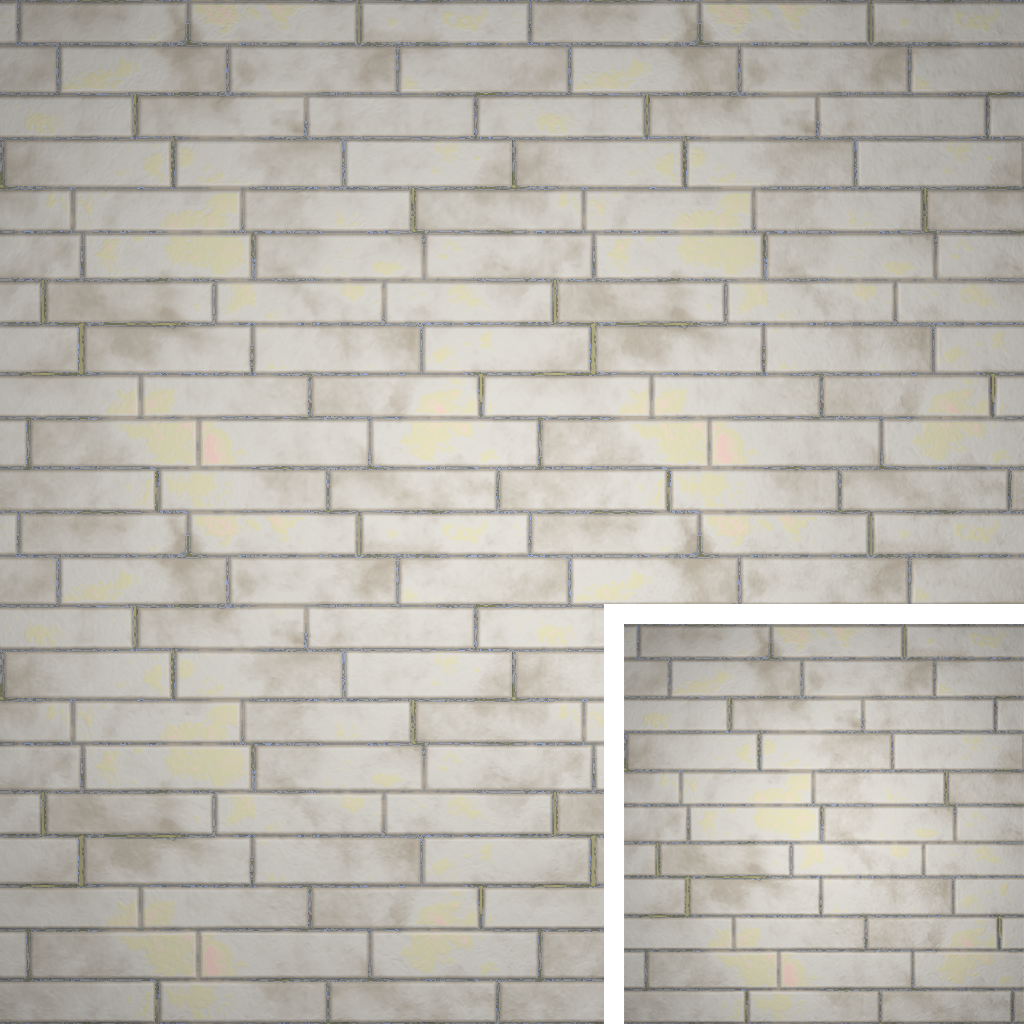}
		} \\
		\multicolumn{4}{c}{Tileable Materials} \\
	\end{tabular}
\caption{Our results are procedural and can easily be edited (First row). After optimization, our results can be synthesized in arbitrary resolution (cropped 2K, cropped optimization target image as inset, 2nd row). Finally, we preserve the generators' tileability, showing here 2x2 tiled results.}
\label{fig:application}
\Description{Applications}
\end{figure}

\section{Results and Comparisons}
In this section, we show our end-to-end material optimization results and compare to previous work. We use our differentiable proxies to enable material graph optimization in two frameworks~\cite{Shi20, hu2022}. We show our material graph optimization results in the MATch framework, using the MATch graph selection step and compare to it in Figs.~\ref{fig:newTeaser} \& \ref{fig:syn_results}. Our approach better matches the target appearance thanks to our optimization of the material structure and scale. We show results with synthetic data in Fig. \ref{fig:syn_results} and with real-world photographs in Fig.~\ref{fig:newTeaser}, showing that our approach can match a variety of generators and appearances. Please see our supplemental materials for additional results. In a quantitative evaluation, the average of feature/style loss of all materials we optimized is 0.408/0.261 for ours against 0.548/1.038 for MATch.

In Fig. \ref{fig:pptbf}, we demonstrate the generality of our differentiable proxy in another inverse modeling framework by Hu et al. \shortcite{hu2022} where we train a PPTBF~\cite{Guehl20} proxy which can simply be plugged into the proposed pipeline to replace their time-consuming structure matching process. Following their approach, we optimize our proxy towards a user-segmented mask map. Our proxy enables gradient-based optimization, reaching similar material appearance to their method, with a 40x speedup (their optimization requires 20 minutes, while ours converges in 30 seconds). Please see our supplemental materials for additional examples.

We highlight that despite the optimization running on fixed resolution, we recover procedural model properties, making our results entirely procedural. Our results can therefore be generated with arbitrary resolution, without a costly high-resolution optimization, and preserve the editing possibilities inherent to procedural models, allowing artists to use our results as a base to kickstart their final vision. We also preserve the original generator properties, such as tileability for the Substance generators, making our results tileable. We demonstrate editability, high resolution material synthesis, and tileability in Fig. \ref{fig:application}.

\begin{figure}
	\centering
	\addtolength{\tabcolsep}{-4pt}
	\begin{tabular}{cccc}
	    Initialization & Ours & Target \\
		\includegraphics[width=0.155\textwidth]{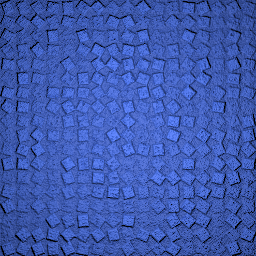} &
		\includegraphics[width=0.155\textwidth]{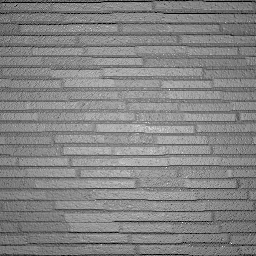} &
		\includegraphics[width=0.155\textwidth]{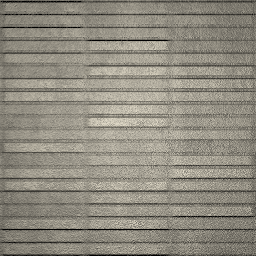} \\
		\includegraphics[width=0.155\textwidth]{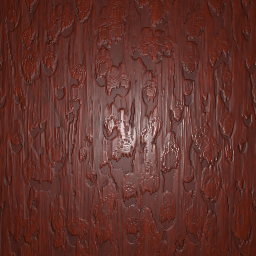} &
		\includegraphics[width=0.155\textwidth]{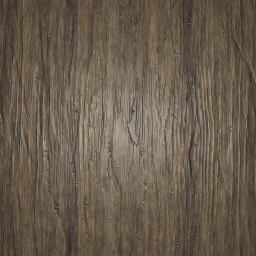} &
		\includegraphics[width=0.155\textwidth]{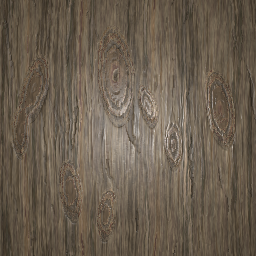} \\ 
		\includegraphics[width=0.155\textwidth]{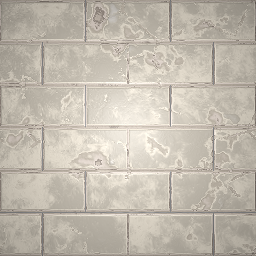} &
		\includegraphics[width=0.155\textwidth]{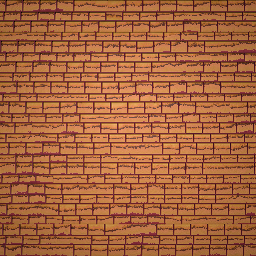} &
		\includegraphics[width=0.155\textwidth]{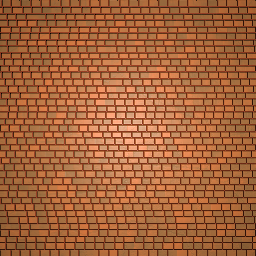} \\
		\includegraphics[width=0.155\textwidth]{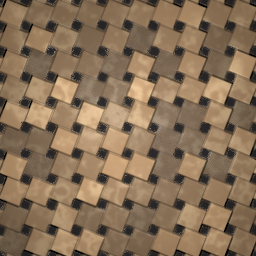} &
		\includegraphics[width=0.155\textwidth]{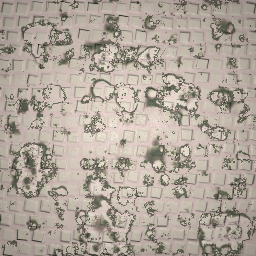} &
		\includegraphics[width=0.155\textwidth]{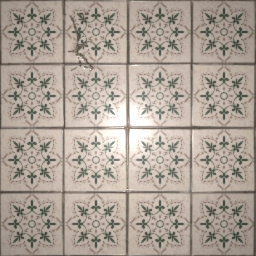} \\
	\end{tabular}
\caption{Limitations. First row: On complex material graphs, in a few cases, our method can lead to a local minimum, despite our initialization approach. Second row: our loss function targets global appearance matching, here it fails to reproduce the knots details in the wood plank. Third row: our proxy fails to reproduce patterns when parameters fall outside of the range used by artists. Here, the number of bricks are beyond our defined sample ranges. Last row: a common limitation in procedural modeling, optimizing a less expressive material graph which does not model specific targeted patterns cannot represent well the target.}
\label{fig:failure_cases}
\Description{Failure Cases}
\end{figure}
\section{Limitations}
We show limitations of our method in Fig.~\ref{fig:failure_cases}. In a Substance we experimented with, the graph structure was such that the optimization was more prone to local minima. This could however be solved by simple manual edits of the graph to simplify the gradient flow.

Another limitation comes from our loss which does not match elements per pixel, missing local details (see wood node in the second row of Fig.~\ref{fig:failure_cases}), or not precisely match the scale (see the first result in Fig.~\ref{fig:application} which has a 7x7 tiles target, but a 6x7 tiles result).

Additionally, due to the importance-sampled parameter space, patterns out of the parameter range used by the artist covered in the training of a proxy are not well reproduced. For instance, when the number of bricks is outside of the distribution, the differentiable proxy cannot reproduce the desired patterns (third row).

Similar to recent inverse procedural material modeling methods \cite{hu2019, Shi20}, our approach fails if the selected initial graph is not expressive enough to match the target appearance, which is illustrated in the last row of Fig.~\ref{fig:failure_cases}.

Finally, although most filters in Substance are differentiable, a few complex filter nodes like FX-map can also have discrete and random behaviors that remain difficult to differentiate. Creating an image-conditioned differentiable proxy for these filter nodes is an interesting future challenge.

\section{Conclusion}
We present a general differentiable solution that optimizes both generator and filter nodes in a material graph. We introduce the Differentiable Proxy, a neural-network-based universal approximator to establish a differentiable parameter space for non-differentiable generators. We demonstrate, with a multi-stage optimization pipeline, that the proxies     enable end-to-end optimization of both structures and appearance to match material photographs or SVBRDF maps. We apply gradient-based optimization, supported by auto-differentiation, on a wide range of material graphs, showing that our framework can achieve high-quality procedural materials from various exemplars. We believe our proxies will enable better material proceduralization and improve differentiability of complex black box functions.

\begin{acks}
This work was supported in part by NSF Grant No. IIS-2007283.
\end{acks}

\bibliographystyle{ACM-Reference-Format}
\bibliography{bibliography} 

\end{document}